# Efficiency Analysis of Materialized views in Data Warehouse Using

# Self-maintenance

Mehwish Aziz / Student at
Govt Post Graduate College, S/town, Rawalpindi
mehwish_aziz90@hotmail.com

Shabnam Nawaz / Student at
Govt Post Graduate College, S/town, Rawalpindi
shabnamgul999@yahoo.com

Pakeeza Batool / Lecturer  at
Govt Post Graduate College, S/town, Rawalpindi
Department of Information Technology
pakeeza_fjwu@yahoo.com

*Abstract*— **A data warehouse is a large data repository for the purpose of analysis and decision making in organizations. To improve the query performance and to get fast access to the data, data is stored as materialized views (MV) in the data warehouse. When data at source gets updated, the materialized views also need to be updated. In this paper, we focus on the problem of maintenance of these materialized views and address the issue of finding such auxiliary views (AV) that together with the materialized views make the data self-maintainable and take minimal space. We propose an algorithm that uses key and referential constraints which reduces the total number of tuples in auxiliary views and uses idea of information sharing between these auxiliary views to further reduce number of auxiliary views.**

I. INTRODUCTION

The problem of materialized views maintenance is very important because views have wide application in data warehousing. A view is derived relation in terms of base relation [5]. View is said to be materialized when the tuples of views are stored in the data warehouse and results are shown without recomputation of view.

When an update occurs and the data at the source changes, the materialized view needs to be updated accordingly so that data at materialized view may be consistent with the data at source, this is called maintenance of materialized views.

There are two approaches to update materialized views when changes occur at the data source. First, we access the source data to identify the change and then recompute materialized views according to that change. Problem with first approach is that it may be too expensive to access the data from the data sources or may be data source is un-available at that time. Second approach to maintenance is usage of self-maintainability. We define self- maintainability as when there is change at the data source, materialized views will be updated using only materialized views and the update. One approach to self- maintainability is to replicate all base data at data warehouse but this approach requires very large storage space and maintenance cost.

Another approach to self- maintainability is to design and place some additional data at the data warehouse. This additional data can be in the form of auxiliary views. Now, challenge is to find most economical auxiliary views in terms of space complexity so that minimal auxiliary views are required such that together MV and AVs are self-maintainable.

*A. Related Work*

Problem of self-maintenance of materialized view has been discussed in [1], [2], [3] and [4]. Inconsistencies occur at the data warehouse because changes at the data source are dynamic. To avoid these inconsistencies, materialized views should be self-maintainable. Self-maintenance can be achieved either by replicating all base data entirely at the data warehouse or by maintaining auxiliary views. Minimizing space and cost of AV is still a research issue [1]. There are many other research areas which still need the attention of the researchers such as update filtering, self-maintainability and multiple view optimization. These problems need to be solved for flexible, powerful and efficient data warehousing. We can use certain shared sub-views to get efficiency but it should be balanced with slow-query response as there are fewer views since some view may not be fully materialized [2]. An algorithm proposed in [3] used idea of information sharing between different auxiliary views so that the number of auxiliary views can be minimal. We improved this approach in our research by using key and referential constraints which further reduces the number of tuples in auxiliary views.

Key and referential integrity constraints are helpful to minimize the number of tuples of auxiliary views discussed in [4].

*B. Our Contribution*

Algorithm discussed in [3] finds minimum number of auxiliary views in AV set. We improve this approach by considering key and referential constraints. When key and referential constraints are used, total number of tuples in AVs will be reduced which further reduces total space occupied by AVs (detail described in section 2.2).

Algorithm discussed in [3] finds set of auxiliary views only through local selection conditions (detail in section 2.1), which are enough to maintain the materialized views but this approach does not take key or referential constraints into considerations. We will find auxiliary views by using RI constraints described in [4] to further reduce the size of auxiliary views in terms of tuples (detail in section 3)

*C. Paper Outline*

Section 2 gives preliminary information and assumptions. Section 3 presents the algorithm that finds minimal auxiliary views set with minimal number of tuples and that set is sufficient for self-maintainability of views. Conclusion is given in section 4.

## II. PRELIMINARIES

*A. Local Selection Condition*

Local selection conditions are those conditions which include attributes from a single relation as opposed to those involving conditions from different relations. When local reduction rule is applied on $R_i$ by pushing local selection conditions, results in views are significantly reduced because those tuples that does not pass local selection condition in $R_i$, does not contribute into views [3].

*B. Finding Auxiliary Views By Considering Key and Referential Integrity Constraints*

We present an example here that will show how we can reduce the number of tuples for auxiliary views needed for the self-maintainability of the views.

We are considering a database of students that has four base relations:-

Department (<u>Dep_no</u>, Dep_name, HOD_name )
Student      (<u>Roll_no</u>, Name, CNIC, FSc_Marks, Dep_no)
Courses     (<u>Course_code</u>, Course_name, Session, Dep_no)
Results       (<u>Result_id</u>, Roll_no, Course_code, GPA)

Each relation has one attribute as a primary key that uniquely identifies records in the relation. In addition, some of these relations have foreign keys which are referenced by another table's primary key. Following Referential integrity constraints holds :

From Student.Dep_no to Department.Dep_no
From Course.Dep_no to Department.Dep_no
From Result.Roll_no to Student.Roll_no
From Result.Course_code to Course.Course_code

Suppose that we maintain a view which contains "Results of students from department of IT, whose session is 2010-2014 along with their name, roll no, course code, GPA and their HOD name".

CREATE VIEW results_IT AS
SELECT   Department.Dep_no,   Department.HOD_name, Student.name,   Student.Roll_no,   Course.Course_code, Student.GPA , Course.Course_name,
FROM Department, Student, Courses, Results
        WHERE Result.Course_code= Course.Course_code
                                and
     Course.Dep_no = Department.Dep_no and
     Result.Roll_no = Student.Roll_no       and
     Student.Dep_no= Department.Dep_no  and
     Course.session = 2010-2014              and
     Department.Dep_name = 'IT'

If we have a view such as above, we have to find such auxiliary views that make data warehouse self-maintainable when insertions are made at base tables. Following is the description of these auxiliary views:

---

CREATE VIEW aux_dept1 AS
SELECT Dep_no, HOD_name
FROM Departmet
WHERE Dep_name='IT';

---

CREATE VIEW aux_std1 AS
SELECT Roll_no, name
FROM Student
WHERE Dep_no IN(SELECT Dep_no FROM aux_dept1)

---

CREATE VIEW aux_course1 AS
SELECT Course_code, Course_name
FROM Courses
WHERE session = 2010-2014 and Dep_no IN (SELECT Dep_no FROM aux_dept1)

---

**Figure1:** Auxiliary views for maintaining the results_IT   view

Figure 1 shows three auxiliary views aux_dept1, aux_std1 and aux_course1 which have been derived from view results_IT. Purpose of these auxiliary views is to maintain the view results_IT when there are some insertions at the source tables. Referential integrity constraints on base relations assure that these three auxiliary views are adequate to maintain results_IT view. There are four base relations in Table1, which contain the number of tuples listed in column 1. Assuming that the selectivity of the Department.Dep_name =0.04 and selectivity of the Course.session=0.05 and that distribution is unvarying. Second column of the Table 1 represents those tuples that pass local selection conditions (section 2.1).We took referential integrity constraints into considerations to further improve the results. Third column of Table 1 shows the results of our approach. According to our example, we did not materialize any tuple from Results table because RI constraints assure that existing tuples in Result cannot join with insertions in other relations due to which total number of AVs will also be reduced.

When we materialized these auxiliary views, it presented the significant saving of space over base relations. As illustrated in Table1.

**Table 1:** Number of tuples for results_IT view

| Base Relation | Tuple in Base Relation | Tuples passing local selection condition | Tuples in auxiliary views |
|---|---|---|---|
| Department | 25 | 1 | 1 |
| Student | 3000 | 3000 | 120 |
| Courses | 1000 | 50 | 2 |
| Result | 1500 | 1500 | 0 |
| Total | 5525 | 4551 | 123 |

These results are shown graphically below:-

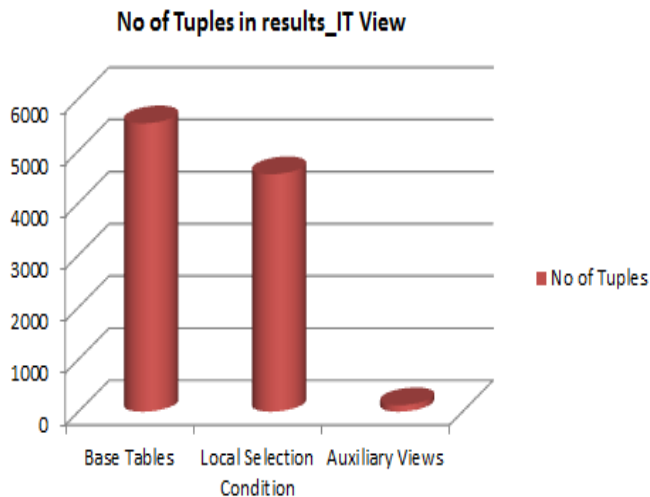

**Figure 2:** Number of tuples for results_IT view

Now we describe the second view result_BBA which will maintain the results of BBA Department i.e Results of students from department of BBA, whose session is 2010-2014 along with their department name, department number, student name, roll no, FSc marks, course code, GPA and HOD name.

CREATE VIEW results_BBA AS
SELECT Department.Dep_no, Department.HOD_name, Department.Dep_name, Student.name, Student.Roll_no, Student.FSc_marks, Student.Course_code, Student.GPA
FROM Department, Student, Courses, Results
WHERE
    Result.Course_code = Course.Course_code and
    Course.Dep_no = Department.Dep_no    and
    Result.Roll_no = Student.Roll_no    and
    Student.Dep_no = Department.Dep_no    and
    Course.session = 2010-2014    and
    Department.Dep_name = 'BBA'

```
CREATE VIEW aux_dept2 AS
SELECT Dep_no, HOD_name
FROM Departmet
WHERE Dep_name='BBA';
```

```
CREATE VIEW aux_std2 AS
SELECT Roll_no, name
FROM Student
WHERE Dep_no IN(SELECT Dep_no FROM aux_dept2)
```

```
CREATE VIEW aux_course2 AS
SELECT Course_code, Course_name
FROM Courses
WHERE session = 2010-2014 and Dep_no IN (SELECT Dep_no FROM aux_dept2)
```

**Figure 3:** Auxiliary views for maintaining the results_BBA view

Figure 3 shows three auxiliary views derived from the view results_BBA. According to our example, three auxiliary views are sufficient to maintain results_BBA view. When we materialized the auxiliary views, it presents the significant saving over base relations. Referential constraints on base relations assure that these three auxiliary views are adequate to maintain results_BBA view. As illustrated in Table 2

**Table 2:** Number of tuples for results_BBA view

| Base relation | Tuple in Base Relation | Tuple passing local selection condition | Tuple in auxiliary views |
|---|---|---|---|
| Department | 25 | 1 | 1 |
| Student | 3000 | 3000 | 120 |
| Courses | 1000 | 100 | 4 |
| Result | 1500 | 1500 | 0 |
| Total | 5525 | 4551 | 125 |

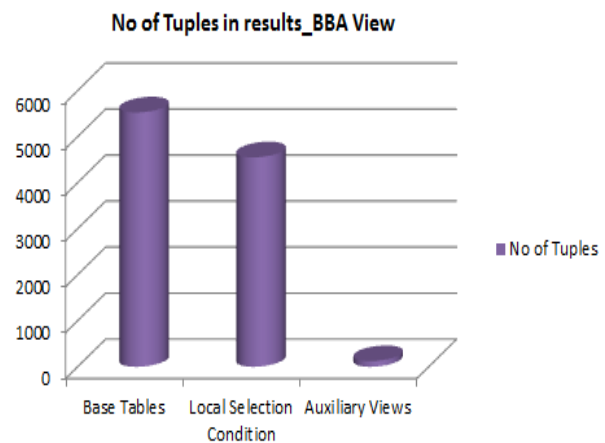

**Figure 4:** Number of tuples for results_BBA view

*C. Gluing Operation*

Gluing operation is applied on two views in order to merge them in to single view. This operation was described in [3]. Here we give only basic overview of this operation. Suppose, there is a relation R, let A and B be the two subsets of R (there may or may not some common attributes in the A and B). Let V1 and V2 be the two views derived from R such that V1=$\pi A^\sigma$ C1and V2= $\pi A^\sigma$ C2 .the gluing operation will find a view V12 that is also derived from R .Our former views V1 and V2 can be derived from our new view V12. Clearly, V12= $\pi AUB^\sigma$ C12 where C12=C1∨C2.

III. PROPOSED ALGORITHM

Input: A set of two materialized views called *'V'*
Output: A set of auxiliary views *'A'* that has minimal number of tuples in auxiliary views.
1. First develop two separate AV sets A1 $\{A^1_{R1}, A^1_{R2},... A^1_{Rn}\}$ and A2 $\{A^2_{R1}, A^2_{R2},... A^2_{Rn}\}$ for two views V1 and V2 by taking key and referential constraints into consideration described in section 2.2.
2. *A:*{}initial AV set
3. Suppose total number of relations in V1 and V2 aren for i=1 to n
   Let $A_{Ri}$ be the resulting view by joining $A_{Ri}^1$ and $A_{Ri}^2$ through gluing operation..
   let $C_i$ be the number of tuples of $A_{Ri}$
   let $n_{ij}$ and $b_{ij}$ be the number of tuples and bytes per tuple in $A^J_{Ri}$
   let $B_i$ be the total number of bytes per tuple in B∩C
   Note: B and C are the subset of attributes of and $A_{Ri}^1$ and $A_{Ri}^2$
   If $C_i(b_{ij}+b_{ij}-B_i) \leq (n_{i1}b_{i1}+n_{i2}b_{i2})$
   then
   $A:= AU(A_{Ri})$
   else
   $A:=\{A_{Ri}^1, A_{Ri}^2\}$
   end if
   end for

*A. How Algorithm Works*

We take our above example to show that how algorithm works. We have two views, one for maintaining the results of IT students along with some of their basic information and second view is for maintaining the results of BBA students.
Both views have their own set of auxiliary views for maintaining views, which we have found using key and referential constraints in first step of our algorithm.
We assume that initial auxiliary view set is empty and we have three auxiliary views in each auxiliary view set. AVs from both sets will be merged in to one AV by using gluing operation. If size of resulting AV is smaller (in terms of tuples and bytes) than two separate AVs ,then it will be included in AV set otherwise two separate AVs will be included in AV set . This algorithm will output a new auxiliary view set which will significantly take less space.
Now we consider an example (section 2.2) to describe our final results.
We have two AV sets.

AV1 {aux_dept1, aux_std1, aux_course1}
AV2 {aux_dept2, aux_std2, aux_course2}

After applying algorithm, we have

AV12{aux_Dep12, aux_Std12, aux_Course12}
Total number of tuples in these AV12 are listed below

**TABLE 3:** NUMBER OF TUPLES WITHOUT REFERENTIAL CONSTRAINTS

| aux_Dep12 | aux_Std12 | aux_Course12 | Total |
|---|---|---|---|
| 2 | 6000 | 200 | 6202 |

Table 3 shows the results of algorithm when we did not consider referential constraints, and when we took referential constraints into account, the number of tuples in auxiliary views are significantly reduced as we can clearly see in table 4.

**TABLE 4:** NUMBER OF TUPLES WITH REFERENTIAL CONSTRAINTS

| aux_Dep12 | aux_Std12 | aux_Cours12 | Total |
|---|---|---|---|
| 2 | 240 | 8 | 250 |

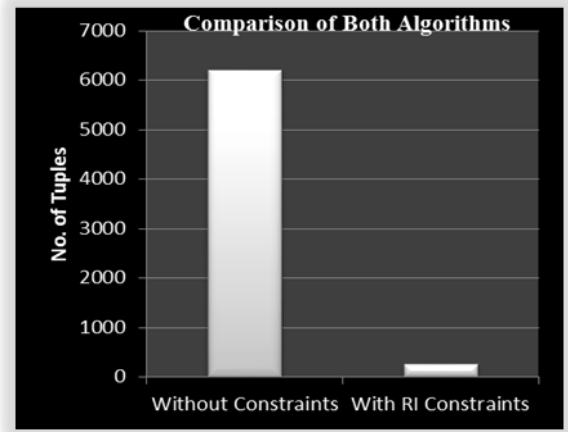

**Figure 5:** Number of tuples in auxiliary view set without and with using RI constraints

Figure 5 shows the results of applying algorithm to results_BBA and results_IT view. When we did not use referential constraints in finding auxiliary view set, total number of tuples are above 6000 and when we used referential constraints we have 250 tuples.

1V. CONCLUSION AND FUTURE WORK

We studied the issue of self-maintainability in this research and devised an algorithm that takes RI constraints into consideration and returns the set of auxiliary view which have minimal number of tuples. In our future work, we will consider effect of key and referential constraints upon deletions at the data sources.